\documentclass[twocolumn,aps,amsmath,amsfonts,amssymb,showpacs,pra,floatfix]{revtex4-1}

\usepackage{color}
\usepackage[dvipsnames]{xcolor}

\usepackage{braket}
\usepackage{amssymb}
\usepackage{bbold}
\usepackage{amsmath}
\usepackage{dsfont}

\usepackage{graphicx}
\usepackage{tikz}
\usepackage[percent]{overpic}
\usepackage{epstopdf}
\usepackage{comment}

\begin{document}
\title{Driven dissipative preparation of few-body Laughlin states of Rydberg polaritons in twisted cavities}

\author{Kristina R. Colladay}
\author{Erich J. Mueller}
\affiliation{Laboratory of Atomic and Solid State Physics, Cornell University, Ithaca, New York 14853, USA}%
\date{\today}

\begin{abstract}
We present a driven dissipative protocol for creating an optical analog of the Laughlin state in a system of Rydberg polaritons in a twisted optical cavity.  We envision resonantly driving the system into a 4-polariton state by injecting photons in carefully selected modes.  The dissipative nature of the polariton-polariton interactions leads to a decay into a two-polariton analog of the Laughlin state.  Generalizations of this technique could be used to explore fractional statistics and anyon based quantum information processing.  We also model recent experiments that attempt to coherently drive into this same state.
\end{abstract}

\maketitle 

\section{Introduction}

Efforts to produce analogs of fractional quantum Hall effects \cite{Stormer1999} in atomic  \cite{Bloch2008,Bloch2012,Wilkin2000,Paredes2003, Wilkin2000, Paredes2001, Cooper2008, Barberan2006, Cooper2001, Sorensen2005, Hafezi2007, Zhang2016, Cooper2013, Roncaglia2011, gemelke2010rotating} and optical settings \cite{Ozawa2019, Maghrebi2015, Anderson2016, Hafezi2013, Chang2008, Birnbaum2005, Wu2017, Schine2016} are driven by the desire to 
directly observe and manipulate ``anyon" excitations, which act as particles with unconventional quantum statistics \cite{Arovas1984, Stern2008, Camino2005, Munoz2020}.   In addition to being of fundamental interest, these anyons can be useful for quantum information processing \cite{Kitaev2003, Lachezar2017, Kapit2012, Kapit2014}.  Here we give a driven dissipative protocol for producing a minimal quantum Hall state in a system of Rydberg polaritons in a non-coplaner optical cavity.

The quantum Hall effect is seen in 2D electron systems in large magnetic fields.  As argued by Schine et al. \cite{Schine2016}, there is a one-to-one correspondence between quantum state of such electrons and the optical modes of a non-coplanar (or twisted) cavity.  This correspondence has lead to a number of proposals to use twisted cavities to reproduce quantum Hall physics \cite{Maghrebi2015,Dutta2018}.  In order to produce an analog of electron-electron interactions, these proposals hybridize the photon modes in the cavity with long-lived Rydberg excitations of an atomic gas.  The resulting Rydberg polaritons interact via a strong dipole-dipole interaction, which 
can be modelled as a short-range repulsion \cite{Beguin2013}.  At particular filling factors, the ground state is expected to be a fractional quantum Hall state.

A key difference between the electronic and optical system is that polariton number is not conserved:  One readily injects polaritons into the cavity by shining an appropriately tuned laser on the system; Polaritons decay due to the finite cavity lifetime, or in inelastic scattering processes.  Here we take advantage of this feature, and show that one can drive this system in such a way that it naturally evolves into an analog of the Laughlin state, one of the most iconic examples of fractional quantum Hall states \cite{Laughlin1983}.

Our scheme is in the spirit of other driven-dissipative state preparation protocols \cite{diehl,PhysRevA.78.042307,vers09,pwave,cian19,sharma2021drivendissipative,budich,bardyn,reiter,Umucahlar2012}, which have analogs in autonomous error correction \cite{Albert2019, Freeman2017, Gertler2021, Combes2021, Liu2016, Cohen2014, Gau2020, Gau2020a}.  The basic idea is that you want to construct a situation where the state of interest is a ``dark state" which is neither excited by the drive, nor decays through the dissipative channel.  We rely upon one of the key features of the Laughlin state, namely that it vanishes when two electrons come close together.  In the Rydberg polariton context, this means that  inelastic two-body loss is strongly suppressed in the Laughlin state.  We construct a drive which takes the system to an excited state, which then decays into the Laughlin state.

In section~\ref{sec: modeling} we present our model of Rydberg Polaritons in twisted cavities.  
We analyze the dynamics via a Lindblad equation, and in section~\ref{sec: numerics} we describe our numerical approach.
In section~\ref{sec: validation} we validate our model by reproducing the observations from \cite{Clark2020}, including calculating the pair correlations functions of emitted photons, one of the key signatures used in the experiment. Section~\ref{sec: interactions} uses our validated model to introduce our driven dissipative protocol for producing the Laughlin state and provides our numerical study of the dynamics.  Section~\ref{sec: measurement} briefly describes how one can verify that that our approach is successful.

\section{Modeling}
\label{sec: modeling}
Researchers in Chicago have designed an optical cavity where the modes are in one-to-one correspondence to the quantum states of 2D harmonically trapped charged particle in a magnetic field.  
 
An atomic gas is placed in the cavity, and the cavity modes hybridize with a long-lived Rydberg excitation.  An idealized effective model \cite{Dutta2018}
for the resulting polaritons, in the cavity's image plane, is $H=H_0+H_{\rm int}$ with
\begin{eqnarray}\label{eqn: single-particle H}
H_0=&\int d^2 r  \frac{1}{2m_0}\hat \psi^\dagger \left(\frac{1}{i}\nabla - e A\right)^2\hat \psi + \frac{1}{2} m_0 \omega^2 r^2 \hat \psi^\dagger \hat{\psi}\nonumber\\
\label{hint}
H_{\rm int}=&\int dr\,dr^\prime\,V(r-r^\prime) \hat{\psi}^\dagger(r)\hat{\psi}^\dagger(r^\prime)\hat{\psi}(r^\prime) \hat{\psi}(r).
\label{eqn: interaction Hamiltonian}
\end{eqnarray}
The operator $\hat \psi(r)$ annihilates a polariton at location $r$, defined in the 2D plane.  The effective vector potential $A$ corresponds to a uniform effective magnetic field, and is related to the non-coplanarity of the cavity.  We use the symmetric gauge and take $A=-\frac{B}{2}y \hat{x} + \frac{B}{2}x \hat{y}$. The coefficients $m_0$ and $\omega$ are related to the cavity geometry and the amount of mixing between the photon modes and the Rydberg excitation.  The polariton-polariton interaction $V(r)$ comes from the dipole-dipole interaction between the Rydberg excitations.  It is short-ranged, and as long as the cavity is sufficiently large it can be replaced by a local potential $V(r)\approx U \delta(r)$.  Some technical details of arriving at this description are given in Appendix~\ref{microscopic}.

The modes which diagonalize Eq.~(\ref{eqn: single-particle H}) have frequencies  $\omega_{mn}= (\omega_c + \Omega_0)n+(\omega_c - \Omega_0)m$, where $\Omega_0 = \sqrt{\omega_c^2 + \omega^2}$, and $\omega_c = \frac{eB}{2m_0}$ which is half the cyclotron frequency
\cite{fock,darwin}. When $\omega_c\approx \Omega_0$ (or equivalently $\omega \approx 0$), modes with different $m$ have the same energy. These nearly-degenerate modes make up the lowest Landau level, and have wavefunctions,
\begin{equation}
\phi_m(x,y) = (x+iy)^me^{-(x^2 + y^2)/d^2}
\label{eqn: landau_wf}
\end{equation}  
with normalization, 
\begin{equation}\label{norm}
1/c_m^2=\int |\phi_m(x,y)|^2 dx\,dy=\frac{d^{2(m+1)}\pi m!}{2^{m+1}}.
\end{equation}
The characteristic length is $\frac{1}{d^2}=\frac{1}{2} m_0 \Omega_0$. 

When multiple polaritons are in the cavity, the interaction term, Eq.~(\ref{hint}), breaks this degeneracy.  The preeminent role of the interactions leads to highly non-trivial physics.  The ground state of this model when there is one polariton for every two modes is the $\nu=1/2$ Laughlin state \cite{Laughlin1983, Clark2020}.  It boasts fractionally charged anyonic excitations.  When there is one polariton per mode, and the interactions are appropriate, one instead finds the $\nu=1$ Pfaffian, with non-abelian excitations \cite{regnault}.  The goal of these experiments is to produce such states and demonstrate their exotic properties.

The defining feature of the $\nu=1/2$ state is that it is the highest density state which vanishes whenever two polaritons touch:
\begin{equation}\label{ll}
\Psi_L(z_1,z_2,\cdots z_N)= \prod_{i<j} (z_i-z_j)^2 e^{- \sum_j |z_j|^2/d^2}.
\end{equation}
As a first step, Clark et al. have isolated three of the modes from Eq.~(\ref{eqn: landau_wf}), corresponding to $m=3,6,9$ (all with $n=0$).  This is sufficient to produce a 2-polariton analog of the Laughlin state,
\begin{equation}
\psi_L(z_1,z_2) \propto z_1^3 z_2^3 (z_1^3-z_2^3)^2 e^{-|z_1|^2/4 d^2 - |z_2|^2/4 d^2}.
\end{equation}
Given the three accessible modes, this is the state with the largest number of polaritons that vanishes when two polaritons touch.  It can also be written as 
\begin{equation}
\ket{L} = \frac{\ket{66} - \sqrt{2.1}\ket{39}}{\sqrt{3.1}},
\label{eqn: laughlin photon}
\end{equation}
where $\ket{66}=\hat a_6^\dagger \hat a_6^\dagger |{\rm vac}\rangle/\sqrt{2}$ and
$\ket{39}=\hat a_3^\dagger \hat a_9^\dagger |{\rm vac}\rangle$.  

Our goal is to construct a protocol for producing the two-photon state, $|L\rangle$.  Once this goal is achieved, one can consider generalizations that will lead to  the many-body state in Eq.~(\ref{ll}). 

We project $H_{\rm int}$ into the relevant space, writing
\begin{equation}
\hat{\psi}(x,y) = \sum_m c_m \hat{a}_m \phi_m(x,y)
\label{eqn: polariton annihilator}
\end{equation}
where $c_m$ and $\phi_m$ are given by Eqs.~(\ref{eqn: landau_wf}) and (\ref{norm}).  This equation defines the annihilation operators $\hat a_m$.
In order to model the experiment, we restrict the sum to be over $m=3,6,9$ -- but in a future experiment it could run over the entire lowest Landau level.

By substituting Eq.~(\ref{eqn: polariton annihilator}) into Eq.~(\ref{eqn: interaction Hamiltonian}) and performing the resulting Gaussian integrals
one arrives at

\begin{equation}
H_{int} = 
 U \sum_{m_1m_2m_3m_4} \Lambda^{m_3 m_4}_{m_1 m_2}  \hat{a}_{m_1}^\dagger \hat{a}_{m_2}^\dagger\hat{a}_{m_3}\hat{a}_{m_4} 
\end{equation}
where the coefficient $\Lambda^{m_3 m_4}_{m_1 m_2}$ vanishes unless $m_1+m_2=m_3+m_4$, in which case it is

\begin{align}
\Lambda^{m_3 m_4}_{m_1 m_2}  =  \frac{1}{\pi d^2 2^{m_1 + m_2}} \frac{(m_1 + m_2)!}{(m_1! m_2! m_3! m_4!)^{1/2}}.
\end{align}
Although we do not make use of it, this interaction is separable:  $\Lambda_{m_1m_2}^{m_3m_4}=\lambda_{m_1m_2}\lambda_{m_3m_4}$.
One can readily verify that Eq.~(\ref{eqn: laughlin photon}) is in the null-space of $\hat H_{\rm int}$.

Polaritons can be injected into the cavity by shining 
lasers with appropriately shaped wavefronts onto the cavity mirror.  This can be modelled by
\begin{equation}
\hat H_{\rm drive} = \sum_m \lambda_m e^{i\nu_mt}\hat{a}_m + \mbox{h.c.}.   
\end{equation}
Here $\nu_m$ is the frequency of the laser which couples to mode $m$.  
As long as $\nu_3+\nu_9=2\nu_6$ one can remove all time dependence by shifting to a rotating frame, in which case the single-particle Hamiltonian is $H_1=H_0+H_{\rm drive}$ with

\begin{align}\label{driveham}
H_0 & = -\omega_3\hat{a}_3^\dagger\hat{a}_3  -\omega_6\hat{a}_6^\dagger\hat{a}_6  -\omega_9\hat{a}_9^\dagger\hat{a}_9\\
\nonumber
H_{\rm drive} &=
\lambda_3\left(\hat{a}_3 + \hat{a}_3^\dagger \right) 
 + \lambda_6\left(\hat{a}_6 + \hat{a}_6^\dagger \right)+\lambda_9\left(\hat{a}_9 + \hat{a}_9^\dagger \right),
\end{align}
here $\omega_m$ is the detuning of the drive from mode $m$.

\begin{figure*}
\includegraphics[width=\textwidth]{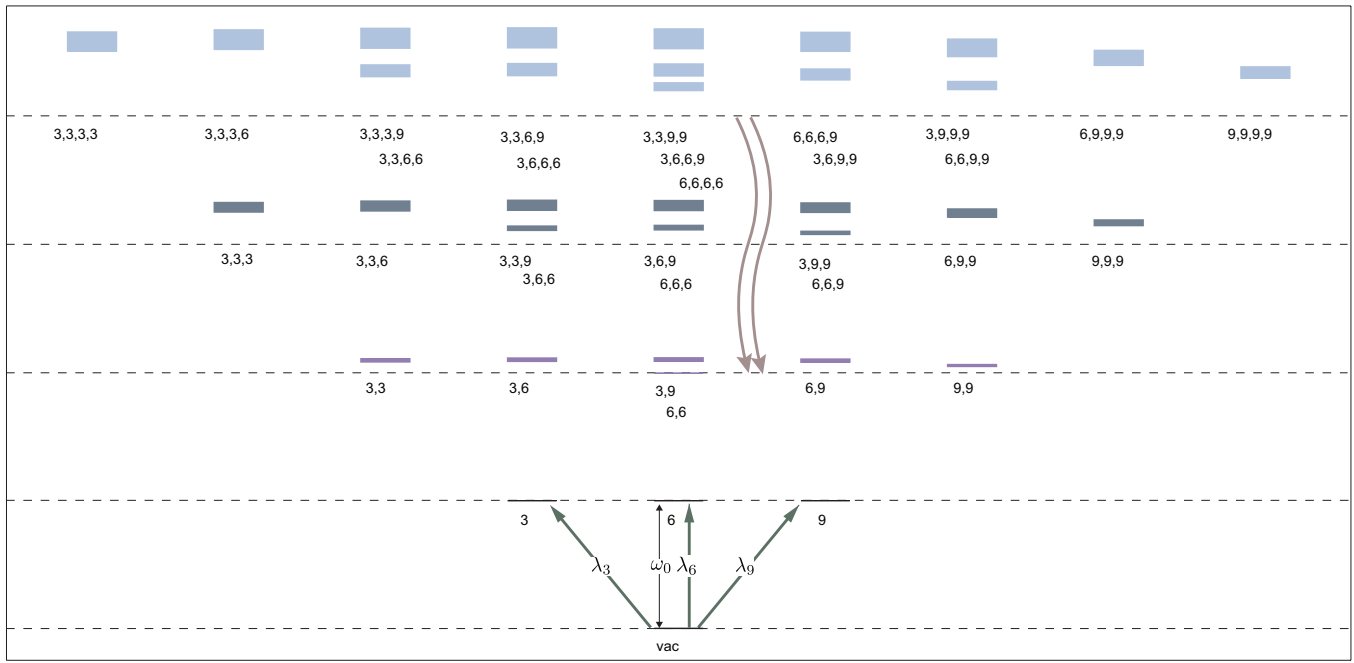}
\caption{(color online) Schematic level diagram for three-mode system ($|3\rangle,|6\rangle,|9\rangle$) with no more than four polaritons.  Vertical axis represents energy, and horizontal axis is the angular momentum relative to a state formed from atoms in the $|6\rangle$ mode: $L-6 N$.  Dashed horizontal lines are separated by $\omega_0$, the mode frequencies in the absence of interactions.  Numbers under each stack of levels represent the angular momenta of the polaritons in the system.  For example, the states labeled with $3,9$ and $6,6$ are superpositions of the states $|3,9\rangle$ and $|6,6\rangle$.  These are split by an energy proportional to the interactions $U$.  The decay rate due to inelastic two-body collisions are represented by the width of each line.  There are four long lived states:  the vacuum, the single polariton states, and the two polariton Laughlin state.  Green arrows are representative single-photon transitions, and brown wavy arrows indicate decay channels from the four photon manifold to the Laughlin state.
}\label{fig: leveldiagram}
\end{figure*}

Polaritons are lost through several processes, which we model using  a Lindblad rate equation for the density matrix $\rho$,
\begin{equation}
\frac{\partial \rho}{\partial t} = -i [H,\rho]+ \left.\frac{\partial \rho}{\partial t}\right|_{\rm incoherent}
\label{eqn: lindblad partial}
\end{equation}
Both the cavity modes and the the Rydberg excitations have finite lifetimes.  These are modelled by the jump operators $\hat L_m=\sqrt{\gamma_1^{(m)}} \hat a_m$.

They contribute
\begin{equation}
\left.\frac{\hat\partial \rho}{\partial t}\right|_{\rm incoherent}^{(1)}=\sum_m D_{\hat L_m,\hat L_m^\dagger}[\hat\rho]
\end{equation}

where
\begin{equation}
D_{\hat A,\hat B}[\hat \rho]= \hat A\, \hat \rho\, \hat B - \frac{1}{2} \hat B \hat A\, \hat \rho - \frac{1}{2} \hat \rho\, \hat B\hat A.
\label{eqn: lindblad operator}
\end{equation}
We treat the single-photon loss rate as independent of the mode index, $\gamma_1^{(m)}=\gamma_1$.  

We also model collisional losses.  As argued in Appendix~\ref{microscopic}, one can engineer the system so that the dominant collisional loss channel involves a process where two proximate polaritons are simultaneously lost.  This is described by
\begin{eqnarray}\label{jump2}
\left.\frac{\partial \rho}{\partial t}\right|_{\rm incoherent}^{(2)}
&=&
\gamma_2\int d^2r D_{\psi(r)\psi(r),\psi^\dagger(r)\psi^\dagger(r)}[\rho]\\
&=& \gamma_2 \sum_j 
\Lambda^{j_3 j_4}_{j_1 j_2}
D_{\hat a_{j_1} \hat a_{j_2},\hat a_{j_3}^\dagger \hat a_{j_4}^\dagger}[\rho].
\end{eqnarray}
Crucially, the decay rate is proportional to the interaction energy, and states with zero interaction energy are infinitely long-lived.  This means that in the absence of single particle loss the Laughlin state has an infinite lifetime.  When restricted to three modes ($m=3,6,9$), the only long-lived states will be the vacuum, the single-polariton states, and the two-photon Laughlin state in Eq.~(\ref{eqn: laughlin photon}).  All others decay due to these two-polariton processes

We will use these incoherent processes as resources to produce the $\nu = 1/2$ Laughlin state by engineering a drive which will excite the system to the four-polariton manifold, relying on loss to populate the long-lived Laughlin state.

Figure~\ref{fig: leveldiagram} shows the energy eigenvalues of $\hat H$, with $4$ or fewer polaritons.   The vertical axis corresponds to energy, and the horizontal axis to angular momentum $L-6N$.  As depicted by the three arrows extending from the vacuum state, the $\lambda_6$-drive couples states which are vertically separated, while the $\lambda_3$ and $\lambda_9$ drives cause diagonal transitions.
The thickness of each energy level corresponds to the decay rate from the collisional two-polariton loss.  

In this diagram, sets of lines have labels which correspond to the angular momenta of the polaritons in the state.  For example, the states above the label 3,9 and 6,6 are the Laughlin state $|L\rangle \propto \ket{66} - \sqrt{2.1}\ket{39}$, and the state $|\bar L\rangle \propto \sqrt{2.1} \ket{66}+\ket{39}$.  As denoted by dashed horizontal lines, in the absence of interactions, all states with $N$ polaritons, would have energy $N\omega_0$, where $\omega_0$ is the bare cavity frequency.  The collisional decay rate is proportional to the interaction energy, so the linewidth grows as one moves away from the dashed line.

\section{Numerical Techniques}\label{sec: numerics}

We use two different numerical techniques to analyze the Lindblad equation.  First, we numerically integrate Eq.~(\ref{eqn: lindblad partial}), using a Runge Kutta finite difference scheme.  This gives us the full time evolution of the $D\times D$ density matrix $\hat \rho$.  In our full model, $D=35$.

Alternatively, we vectorize the Lindblad equation interpreting $\hat \rho$ as a length $35^2$ vector, $\vec \rho$.  The Lindblad equation becomes $\partial \vec \rho/\partial t={\cal L}\rho$, with
\begin{align}
\nonumber
\mathcal{L} & = -i H \otimes \mathds{1} + i \mathds{1} \otimes H^T + \Gamma\sum_\nu L_\nu \otimes(L_\nu^\dagger)^T \\
& \qquad - \frac{1}{2} \left(L_\nu^\dagger L_\nu \otimes \mathds{1} + \mathds{1}\otimes\left(L_\nu^\dagger L_\nu \right)^T\right).
\end{align}  
The steady state corresponds to the kernel of the $35^2 \times 35^2$ matrix $\mathcal{L}$.  We numerically find the nullspace of $\mathcal{L}$ using standard linear algebra libraries.

\section{Validation}\label{sec: validation}
Before implementing our dissipative approach to generating the Laughlin state, we validate parts of our model by  reproducing 
the experimental observations in

\cite{Clark2020}.  In addition to giving us the opportunity to compare to experimental results, the simpler nature of this experiment lets us explicitly write down all expressions, making our arguments more concrete.

\begin{figure}
\centering
\includegraphics[width=.5\columnwidth]{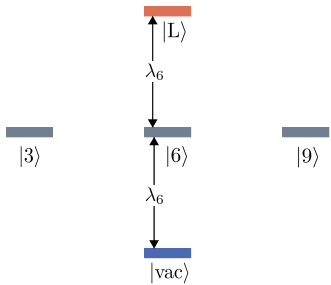}
\caption{(color online) Energy-level diagram depicting the relevant states in the experiment of Clark et al. \cite{Clark2020}, analyzed in Sec.~\ref{sec: validation}. A resonant drive connects $|{\rm vac}\rangle$,  $\ket{6}$ and $\ket{L}$. The modes $\ket{3}$ and $\ket{9}$ can become occupied through single-photon decay from $\ket{L}$.}
\label{fig: direct_drive}
\end{figure}

In \cite{Clark2020} the experimentalists resonantly drive the $|6\rangle$ mode:  $\omega_6=\nu_6-\omega_0=0$.  The other two modes are undriven: $\lambda_3=\lambda_9=0$.  Since the states $|{\rm vac} \rangle$,$|6\rangle$, and $|L\rangle$ are resonantly coupled,  and the drive strengths are low, there is essentially zero probability to end up in any other state except these three, and the states they decay into: $|3\rangle$ and $|9\rangle$.
It therefore suffices to truncate to the simplified model depicted in
Fig. \ref{fig: direct_drive}.
  
In the rotating frame, the Hamiltonian is
\begin{equation}
H = 
\begin{bmatrix}
0 & 0 & \Omega & 0 & 0 \\
0 & 0 & 0 & 0 & 0 \\
\Omega & 0 & 0 & 0 & \Omega' \\
0 & 0 & 0 & 0 & 0 \\
0 & 0 & \Omega' & 0 & 0
\end{bmatrix}
\label{eqn: 5-state_coherent_Hamiltonian}
\end{equation}
where the basis states are
$\ket{\mbox{vac}}, \ket{3}, \ket{6}, \ket{9}, \ket{\mbox{L}}$.
The matrix elements are
\begin{align}
\Omega &= \bra{\mbox{vac}} \lambda \hat{a} \ket{6} = \lambda \\
\Omega'  &= \bra{6} \lambda \hat{a} \ket{L} = \sqrt{\frac{2}{3.1}}\lambda,
\end{align}

\begin{figure}
\includegraphics[width= \columnwidth]{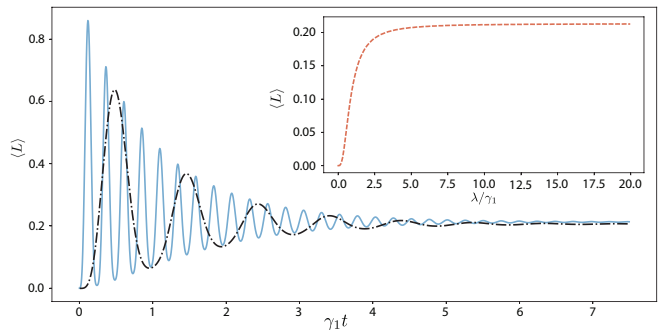}
\caption{(color online) Time evolution of the probability of being in the Laughlin state for the scenario in Fig.~\ref{fig: direct_drive}. Here $\gamma_1$ is the single-polariton loss rate, and $\lambda=\lambda_6$ is the drives strength. Solid blue: $\lambda/\gamma_1 = 20$; Dashed-dotted black: $\lambda/\gamma_1 = 5$.  
Inset: Steady-state probabilities of the Laughlin state as a function of drive strengths.}
\label{fig: drive_hard}
\end{figure}
None of these states experience collisional losses, so we only need to consider the single-photon loss terms, corresponding to jump operators
\begin{align}
\hat L_3 &= \ket{\mbox{vac}}\bra{3} - \frac{\sqrt{2.1}}{\sqrt{3.1}}\ket{9}\bra{\mbox{L}}\\
\hat L_6 &= \ket{\mbox{vac}}\bra{6} + \frac{\sqrt{2}}{\sqrt{3.1}}\ket{6}\bra{\mbox{L}}\\
\hat L_9 &= \ket{\mbox{vac}}\bra{9} - \frac{\sqrt{2.1}}{\sqrt{3.1}}\ket{3}\bra{L}
\end{align}
We numerically integrate the Lindblad equation, 
\begin{equation}
\partial_t \hat\rho = i[\rho,H] + \gamma_1(D_{L_3,L_3^\dagger}\rho + D_{L_6,L_6^\dagger}\rho + D_{L_9,L_9^\dagger}\rho),
\label{eqn: lindblad 5 state}
\end{equation}
Starting from the vacuum state at time $t=0$.
Figure~\ref{fig: drive_hard} shows  typical results for the time evolution of the probablility of being in the Laughlin state $\langle L\rangle= \langle L | \hat \rho |L\rangle$.  

If the drive is weak compared to the single-photon loss rate, $\lambda \ll\gamma_1$, then very few polaritons are ever in the cavity:  As soon as a polariton is created, it decays.  In the opposite limit $\lambda \gg \gamma_1$, damped Rabi oscillations are seen.  As shown in the inset, the  steady-state population of the two-photon Laughlin state is a monotonic function of the drive amplitude, saturating at roughly 21$\%$ when the drive is very strong.  
 \begin{figure}
\includegraphics[width=\columnwidth]{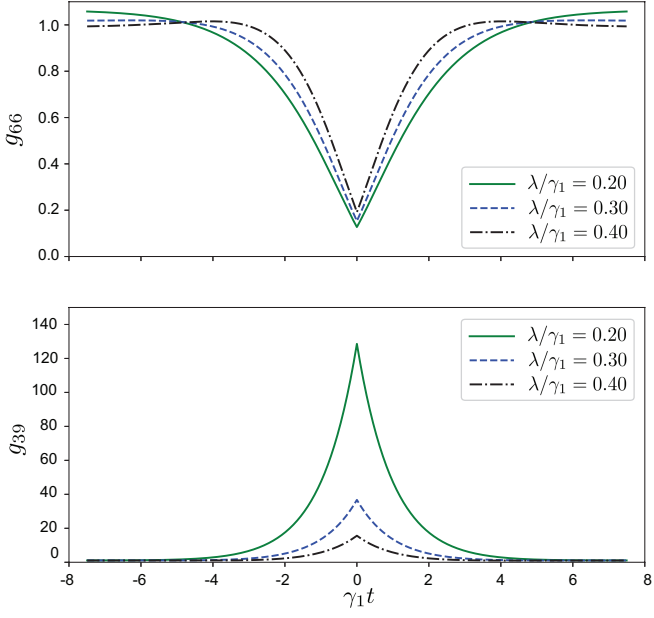}
\caption{(color online) Correlation functions $g_{ij}=c_{ij}/\Gamma_i\Gamma_j$ corresponding to the scenario in  Fig.~\ref{fig: direct_drive}.  
Here $c_{ij}(t)$ is the joint probability of emitting a photon into mode $i$ and a second into $j$, with a time delay of $t$; $\Gamma_i$ is the steady state probability of emitting into state $i$.   
 The single-photon decay rate is $\gamma_1$, and the drive strength is $\lambda$.}
\label{fig: correlation}
\end{figure}

In the experiment \cite{Clark2020}, $\lambda/\gamma_1\sim 0.35$, and one therefore expects that the steady-state ensemble only has a 1.8$\%$ probability of being in the Laughlin state.  To verify that they have produced the Laughlin state, the experimentalists use a correlation measurement:  If two photons are simultaneously emitted from the cavity, then the cavity must have been in a two-polariton state. 

Aspects of the two-photon state can be extracted from the mode-structure of the outcoming photons.  This can be thought of as a form of post-selecting:  ``What is the quantum state when two polaritons are in the cavity?"

More precisely, the experimentalists measure the steady state correlation function
$g_{jk}(\tau)=c_{jk}(\tau)/\Gamma_j \Gamma_k$, where  $c_{jk}(\tau)$ is the joint probability of measuring a photon in mode $j$ at time $t$ and another in mode $k$ at time $t+\tau$.  This quantity is  normalized by the the product of the probabilities of measuring each of these events separately, $\Gamma_j$ and $\Gamma_k$.  In steady-state the denominator is time independent.

Results from our numerical calculations are shown in Fig.~\ref{fig: correlation}.  In the top panel one sees that sequential photons in the 6-mode are anticorrelated.  At delay $\tau=0$, 
\begin{equation}
g_{66}(0)=\frac{
(\gamma_{L\to 6}) (\gamma_{6\to L}) P_L}{\left( (\gamma_{6\to \rm vac})P_6 + ( \gamma_{L\to 6})P_L\right)^2 }=\frac{\frac{1}{3.1}  P_L}{(P_6+\frac{2}{3.1} P_L)^2},
\end{equation}
where $P_6$ and $P_L$ are the steady state probabilities of being in the $|6\rangle$ or $|L\rangle$ states.  The rate of emitting a $6$ photon from $|L\rangle$ is $\gamma_{L\to 6}=2 \gamma_1 /3.1$, while the rate for emitting from $|6\rangle$ is $\gamma_{6\to\rm vac}=\gamma_1$.  
  
For large delay, $g_{66}(\infty)=1$, as the two events are uncorrelated.  
The photons are anticorrelated at short times because once a photon is emitted from the $|6\rangle$ state, a second cannot be emitted until another photon enters the cavity.  

For larger $\lambda/\gamma_1$ one sees weak oscillations in the correlation function, as Rabi oscillations between the states make certain delays less likely than others.

The photons in the $|3\rangle$ and $|9\rangle$ modes are positively correlated.  Whenever a $|3\rangle$ photon is emitted, a $|9\rangle$ photon must follow at a later time.  At zero time delay
\begin{eqnarray}
g_{39}(0)&=& \frac{
(\gamma_{L\to 3})(\gamma_{3\to\rm vac})P_L}{((\gamma_{3\to \rm vac})P_3+\gamma_{L\to 9} P_L)((\gamma_{9\to \rm vac})P_9+\gamma_{L\to 3} P_L)}\nonumber\\&=&\frac{
\frac{2.1}{3.1}  P_L}{(P_3+\frac{2.1}{3.1} P_L)(P_9+\frac{2.1}{3.1} P_L)},\label{g66}
\end{eqnarray}
where, as before,  the rate of emission from $i$ to $j$ is $\gamma_{i\to j}$:  $\gamma_{L\to 3}=\gamma_{L\to 9}=\gamma_1 (2.1/3.1)$ and $\gamma_{3\to vac}=\gamma_{9\to vac}=\gamma_1$.  Equation~(\ref{g66}) can be somewhat simplified by using the principle of balance: $P_L \gamma_{L\to 3}= P_3 \gamma_{3\to vac}$, and  $P_L \gamma_{L\to 9}= P_9\gamma_{9\rightarrow\mbox{vac}}$.  This yields
\begin{equation}\label{g66simp}
g_{39}(0)=\frac{1}{4} \frac{3.1}{2.1} \frac{1}{ P_L}.
\end{equation}
When $P_L$ is small, then the time between emission of the individual photons in a 3-9 pair is short compared to the time between event pairs.  This leads to a large correlation peak.  
At long times $g_{39}(\infty)=1$, as one is detecting photons from different pairs.  The correlation function should just fall exponentially between these values, with decay time $\tau_1$.

To numerically calculate $g_{jk}(\tau)$, we first integrate Eq.~(\ref{eqn: lindblad 5 state}) for a long time to produce the steady state density matrix 
$\hat \rho_\infty$.  
We calculate the emission rates $\Gamma_j = \gamma_1 {\rm Tr} \hat a_j^\dagger \hat a_j \hat\rho_\infty= {\rm Tr} \hat a_j \hat \rho_\infty \hat a_j^\dagger $.  
The density matrix immediately following a $j$-photon emission event is $\hat \rho_j= \hat a_j \hat \rho_\infty \hat a_j^\dagger/\Gamma_j$.  
We then evolve this density matrix for time $\tau$ to calculate $c_{jk}= \Gamma_j {\rm Tr} \hat a_k^\dagger \hat a_k \hat \rho_j(\tau)$, and $g_{jk}=c_{jk}/\Gamma_j\Gamma_k$.

Another important probe in \cite{Clark2020} is to compare the probability of simultaneously observing a
$3$ and $9$ photon, to that of finding 
two $6$ photons, 
\begin{eqnarray}
R&=&\frac{c_{39}(0)+ c_{93}(0)}{c_{66}(0)}.
\end{eqnarray}
Within this truncated model, the only contributions to the correlation functions come from the diagonal element of $\hat \rho$  corresponding to the Laughlin state,
\begin{eqnarray}
R&=& \frac{P_L \langle L | a_3^\dagger  a_9^\dagger a_9 a_3|L\rangle+P_L \langle L | a_9^\dagger  a_3^\dagger a_3 a_9|L\rangle}{P_L \langle L | a_6^\dagger a_6^\dagger a_6 a_6|L\rangle}\\
&=& 2.1.
\end{eqnarray}
The experimental results are within error bars of this number, which is consistent with 
producing the Laughlin state.
 
\section{Driven-Dissipative  Preparation of the 2-polariton Laughlin State}
\label{sec: interactions}
\begin{figure}
\centering
\includegraphics[width= \columnwidth]{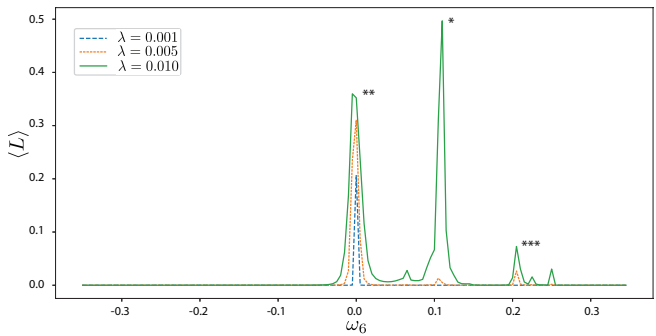}
\caption{(color online) Steady state probabilities of the Laughlin state as a function of the detuning.  Each of the three modes are driven with equal amplitude: $\lambda_3=\lambda_6=\lambda_9=\lambda$. The $m=6$ detuning, $\omega_6$ is displayed on the horizontal axis, and the $m=3$ and $m=9$ drives are chosen such that the four-body state $\ket{A} = -0.362\ket{3339} - 0.932\ket{3366}$ is on resonance.  The energy level diagram corresponding to this drive condition is depicted in Fig. \ref{fig: energy_diagram}.  For reference, the three highest peaks are labeled by different numbers of asterisks.
}
\label{fig: probL_vs_w6}
\end{figure}

As illustrated by the model 
in Sec.~\ref{sec: validation}, in the absence of careful timing, one cannot efficiently produce the Laughlin state by resonant coupling.  This is a generic feature of coherent quantum systems:  For example,  resonantly coupling the levels of a two-level system will not create a steady-state inverted population.  On the other hand, a laser can be made from a 3-level atom.  There one produces a large occupation of an excited metastable state by driving to a third level which decays into it. We will explore the many-body analog of this approach.  Four-polariton states play the analog of the unstable excited level, and the Laughlin state plays the role of the metastable excited level.  Collisional two-polariton loss provides the decay from the four-polariton manifold to the Laughlin state.

The challenge in designing this protocol is that many levels are involved.  Figure~\ref{fig: leveldiagram} shows a complicated network of 35 levels. Each $N$ polariton states couple to  three states with N+1 polaritons. This gives 60 distinct transition matrix elements.  One needs to be able to efficiently couple the vacuum state to the four-photon manifold without coupling the Laughlin state to any other level.  Moreover there are a large number of parameters, including the amplitude and frequencies of the three drives, the interaction strength, and the various decay constants.

The challenge would be greatly reduced if we had a technique to coherently inject 4-photons, directly coupling the vacuum state to the four-polariton manifold.  Appendix~\ref{appendix: multi_photon_drive} analyzes that simplified case.  Here we tackle the full problem, where the drive is of the form in Eq.~(\ref{driveham}).

Exciting the system into the 4-polariton manifold is a fourth order process -- which will be very strongly suppressed when the drive is weak.  One cannot simply increase the drive strength, as a strong drive will also deplete the Laughlin state.  
Our strategy for overcoming this difficulty is to engineer a string of intermediate states.  Ideally the photons would be absorbed through  four resonant (or near resonant) transitions.  Simultaneously we need the Laughlin state to be spectrally isolated.

We adjust the detunings $\omega_3$, $\omega_6$, and $\omega_9$ to optimize this process.  As previously explained, in order to have a well-defined rotating frame, we require $\omega_3+\omega_9=2\omega_6$.  Furthermore, we require that the four-photon transition is resonant.  There are ten 4-polariton states which have a finite probability to decay into the Laughlin state.    
As a concrete example, consider $|A\rangle=-0.362 |3339\rangle-0.932 |3366\rangle$, with interaction energy $E_A=1.245U$.  To resonantly couple to this state, we need $3\omega_3+\omega_9 = 2\omega_3+2\omega_6=E_A$.  Throughout we use units where $U=1$ and take $\gamma_2=0.001$.  We neglect single-polariton loss, taking $\gamma_1=0$.

  \begin{figure}
\begin{overpic}[width=\columnwidth,percent]{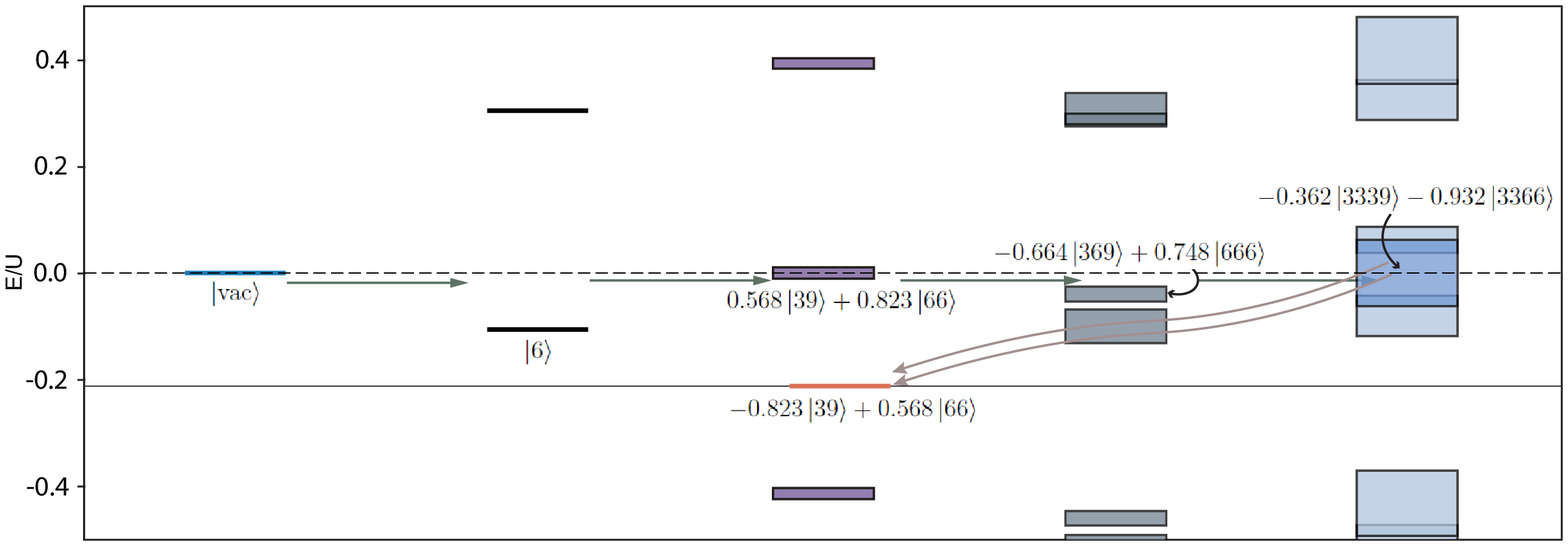}
 \put (15,30) {*}
\end{overpic}
\begin{overpic}[width=\columnwidth]{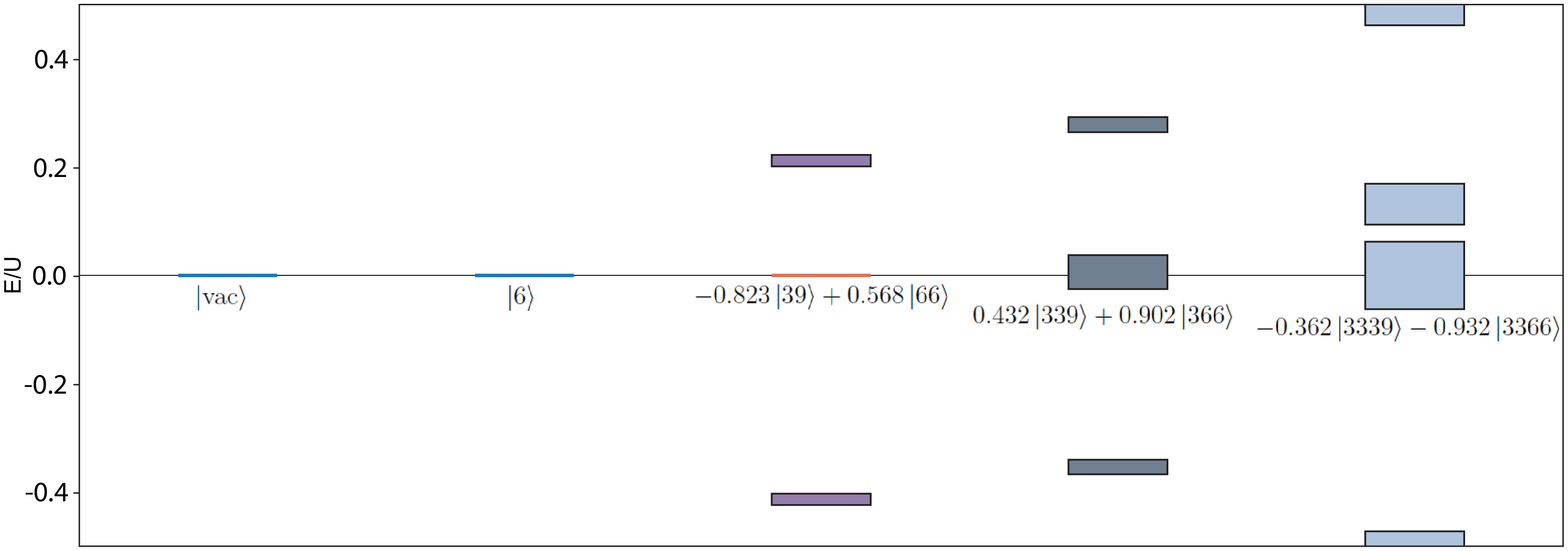}
 \put (15,30) {**}
\end{overpic}
\begin{overpic}[width=\columnwidth]{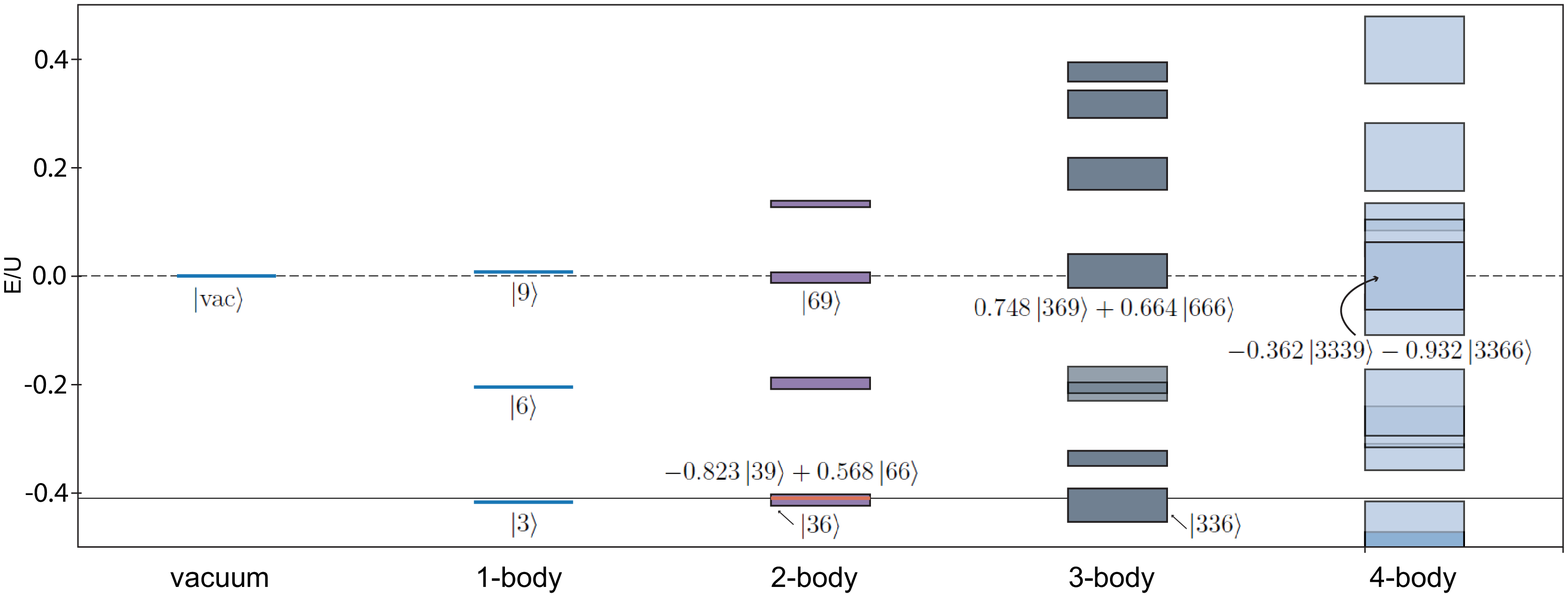}
 \put (15,30) {***}
\end{overpic}
\caption{(color online) Rotating frame energy level diagrams corresponding to the three most prominent peak Fig. \ref{fig: probL_vs_w6}, marked by (*)  top, (**) middle, and (***) bottom.  As illustrated by horizontal arrows in the top panel, a single-photon drive sequentially couples states which differ by one polariton.  The Laughlin state is populated by two-body loss from the four-body manifold (denoted by the double arrow).}
\label{fig: energy_diagram}
\end{figure}

Figure~\ref{fig: probL_vs_w6}
illustrates the role of resonances by plotting
the steady-state probability of being in the Laughlin state as a function of  $\omega_6$, calculated by the technique in Sec.~\ref{sec: numerics}.  Here we fix $\omega_3=E_A/2-\omega_6$, so that the state $|A\rangle$ remains degenerate with the vacuum state, in the rotating frame.  One sees at least six discrete peaks, corresponding to when various intermediate states become resonant.  
    The most prominent peaks correspond to the simultaneous alignment of multiple intermediate states.  Figure~\ref{fig: energy_diagram} shows energy level diagrams
corresponding to the three largest peaks.  These depict the rotating-frame energy of the states from Fig.~\ref{fig: leveldiagram}, separated into columns, each of which represents a different number of polaritons.  As illustrated by the arrows, the drive changes the polariton number by 1, and hence connects states in neighboring columns.  Two-polariton collisional loss causes an incoherent decay into a state which is two columns to the left.  Energy conservation is relevant to the coherent drive processes, but not the loss processes.  As before, the line-widths are shown by the thickness of the lines.  These three energy level diagrams illustrate the central design principles behind our approach.  

The most important feature of the level diagrams in Fig.~\ref{fig: energy_diagram} is the fact that there is a direct chain of near-resonant excitations that takes one to the four-polariton manifold.  The closer these intermediate states are to resonance, the higher the effective transition rate to the four-polariton state.  The second most important feature is how spectrally isolated the Laughlin state is.  The reason that the peak marked (*)  is higher than the one marked (***)  is that the Laughlin state is more isolated.  One is led to the intuitive recipe:  Align the the levels needed to absorb photons, but keep the Laughlin state isolated.

The second highest (**) peak in Fig.~\ref{fig: probL_vs_w6}  is qualitatively different than the others, as the Laughlin state is coherently coupled to the vacuum.  In that sense it is more similar to the scenario in Sec.~\ref{sec: validation}, and it shares the qualitative features of that model.  There are however two differences.  First, in this section we are driving with $\lambda_3$, $\lambda_6$, and $\lambda_9$, while in Sec.~\ref{sec: validation} only $\lambda_6$ was nonzero.  The three drives add coherently, and result in a somewhat larger population of the Laughlin state.  The other difference between this scenario and Sec.~\ref{sec: validation} is that there are 3-polariton and 4-polariton states which are also resonant.  The minimal model here requires including these two additional states, resulting in a 7 dimensional Hilbert space.

\begin{figure}
\begin{overpic}[width=\columnwidth]{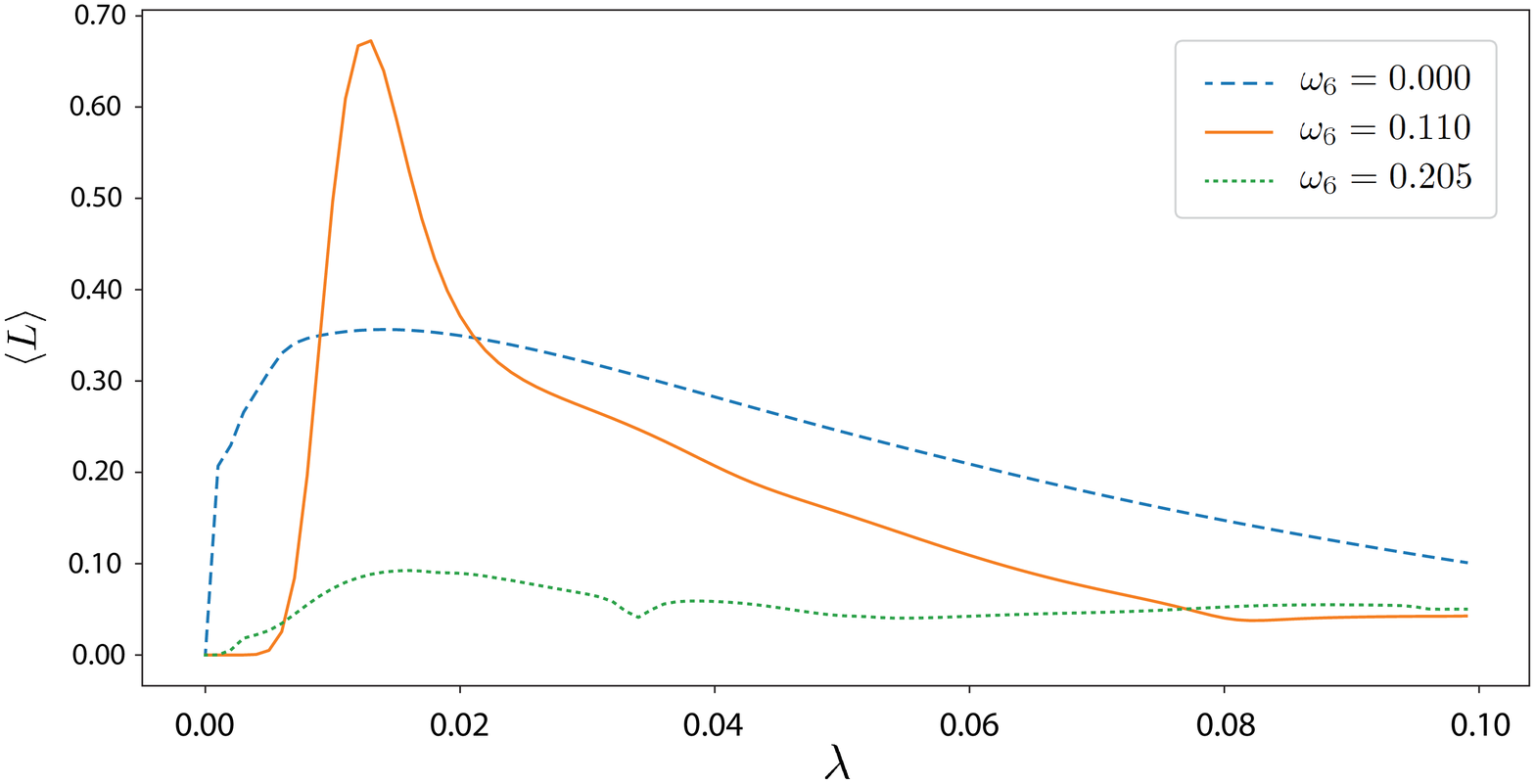}
\put(20,45){*}
\put(80,18){**}
\put(30,10){***}
\end{overpic}
\caption{(color online) Steady state probability of the Laughlin state for the three largest peaks in Fig. \ref{fig: probL_vs_w6} as a function of drive amplitude $\lambda_3=\lambda_6=\lambda_9 = \lambda$ for $\gamma_2 = 0.001$.}\label{fig: amp}
\end{figure}
   
As shown in Fig.~\ref{fig: amp}, the probability of ending up in the Laughlin state is a non-monotonic function of the drive-strength.  For weak drive, the bottle-neck is populating the four-photon manifold.  In this regime, the population of the Laughlin state increases with drive strength.
The rate of increase depends on the accuracy of the alignment of the intermediate states:   Thus for small $\lambda$ the solid red curve (corresponding to the peak with a single asterisk in Fig.~\ref{fig: probL_vs_w6}) rises slower than the dotted green curve (corresponding to three asterisks).  As should be apparent, the physics of the dashed curve is somewhat different, as the Laughlin state is resonantly excited.

If the drive is made too strong, the probability of ending up in the Laughlin state falls.  This occurs because the drive  excites the system out of the Laughlin state. 

\begin{figure}
\includegraphics[width=\columnwidth]{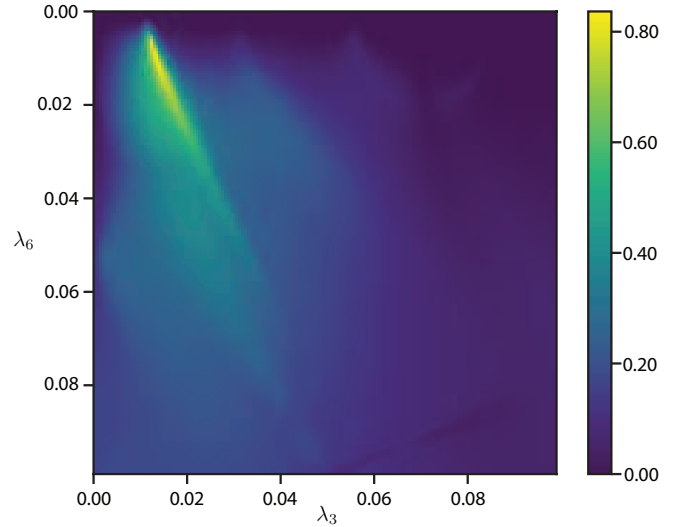}
\caption{(color online) Steady state probability of being in the Laughlin state for detuning $\omega_6=0.1064$ (corresponding to the (*) state in Fig.~\ref{fig: probL_vs_w6} as a function of $\lambda_3$ and $\lambda_6$ drive amplitudes with $\lambda_9=0$.  The peak probability is $83.7\% $ at $\lambda_6 = 0.010$ and $\lambda_3 = 0.013$.}\label{amplitude}
\end{figure}

The story is somewhat complicated by the fact that there are three different $\lambda$'s.  Figure~\ref{amplitude} further explores the amplitude dependence of the (*) peak in Fig.~\ref{fig: probL_vs_w6}.  We set  $\lambda_9=0$ and vary $\lambda_3$ and $\lambda_6$.  The most prominent feature is a peak at $\lambda_6=0.010, \lambda_3=0.013$, and a ridge which extends out from it at an angle.  This ridge is a quantum interference effect:  For a certain ratio of $\lambda_6/\lambda_3$, the  excitations to the four-polariton manifold are enhanced.  Due to the presence of many different intermediate states, and many states which couple to the Laughlin state, the structure can be quite rich:  For example, the dotted green curve in Fig.~\ref{fig: amp} has a dip near $\lambda=0.035$, and the red curve has several kink-like features.
    
We systematically explored all ten potential four-photon transitions, producing the analog of Fig.~\ref{fig: probL_vs_w6} for each of them.  For each of the large peaks we numerically optimized the values of $\lambda_3,\lambda_6,\lambda_9$ and $\omega_6$.  The most favorable example is the one in Fig.~\ref{fig: energy_diagram} which resulted in  a steady-state occupation of the Laughlin state of 83.9\% with $\lambda_6=0.0086, \lambda_3=0.0129$, and $\lambda_9 = 0.0014$.

\begin{figure}
\centering
\includegraphics[width= \columnwidth]{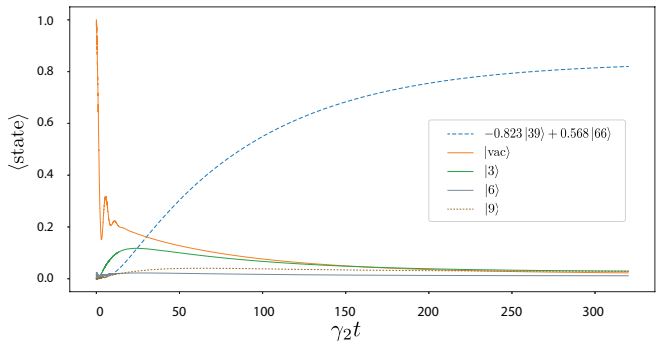}
\caption{(color online) Probability of the long-lived states $\ket{\mbox{vac}}, \ket{3}, \ket{6}, \ket{9}$ and $\ket{L}$ as the system in the (*) configuration evolves according to the Lindblad master equation. The time scale for the finite evolution is in terms of the two-body loss rate. The evolution uses optimized parameters for this configuration. The system configuration at long times contains a non-zero population of the long-lived single-body states.}
\label{fig: evolution}
\end{figure}

\section{Measurements}
\label{sec: measurement} 
The most easily interpreted probe of our system is monitoring the light emitted by the two-polariton loss events.  No such events occur in the Laughlin state, and the disappearance of this signal is a direct measurement of the short range polariton-polariton correlation function. Analogous correlation measurements were carried out in atomic systems \cite{gemelke2010rotating}.  The challenge of this measurement is that the light is presumably emitted in random directions.  Furthermore, this signal cannot distinguish between different dark states.

A slightly more indirect, but simpler measurement is monitoring the reflected light from the drive.  Any photons which enter the cavity show up as an attenuation of this reflected light.  Thus when the system reaches the Laughlin state, which is dark, the reflected intensity increases.  Again, any dark state would give the same signal.

A third approach is to monitor the intensity and correlations of the light emitted through single-polariton losses, as was discussed in Sec.~\ref{sec: validation}.  Since this probe is harder to interpret, it is not as useful as a ``smoking gun."  Its primary advantage is that the measurement is relatively straightforward.

\section{Summary and Discussion}

We presented a driven dissipative approach to producing the two-particle Laughlin state of polaritons in a twisted cavity.   This is part of a broader goal of producing and braiding anyons in an AMO system.

Producing the two-particle Laughlin state is a modest, but necessary step towards this goal.  Our technique is a useful complement to existing proposals, and it brings with it a number of pros and cons.

Most importantly, our technique is self-correcting:  If a perturbation (such as decay of a polariton) knocks the system out of the Laughlin state, the driven-dissipative process will bring you back.  It also does not rely upon any careful timing. 

Unfortunately, our technique does require fine-tuning the of the drive frequencies and amplitudes.  In an experiment, this fine-tuning would need to be done by trial and error, as our models have finite accuracy.  These experiments also require a hierarchy of energies which may be difficult to attain -- namely:  $\gamma_1 \ll \lambda, \gamma_2 \ll U$.  Here $\gamma_1$ is the rate of single-photon loss, $\lambda$ is the drive strength, $\gamma_2$ is the rate of interaction-driven two-photon loss, and $U$ is the interaction strength.  The most challenging aspect, requiring novel cavities, is making $\gamma_1$ smaller than the other scales.  The relative size of $\gamma_2$ and $U$ can be controlled by adjusting the detuning of the EIT transition used for the Rydberg polaritons (Appendix~\ref{microscopic}).  The drive strengths $\lambda$ are under direct experimental control.

A crucial question is the scalability of the approach.  The biggest impediment here is that as one adds more states, the energy-level spectrum becomes quite dense.  Consequently even more fine-tuning would be needed to create larger Laughlin states.  Nonetheless, this driven-dissipative approach is a valuable tool for state creation and stabilization. 

\section{Acknowledgements}
We thank Jon Simon and his group for fruitful discussions.  This work was supported by NSF PHY-1806357.

\appendix

\section{Microscopic Model of interacting Rydberg Polaritons in nearly degenerate multimode cavity}
\label{microscopic}
For completeness we give a brief derivation of the model from Sec.~\ref{sec: modeling}.  We make no attempt at mathematical rigor -- more thorough treatments can be found elsewhere \cite{simontheory,bienias,gullans,gorshkov,maghrebi}.

The modes of an optical cavity are labeled by a longitudinal index $k$, and transverse indices $\alpha$.  In the image plane, where the atoms sit, the electric field has magnitude $E_\alpha^{(k)}(x,y)$.  
We are considering the case where several modes with the same $k$ are nearly degenerate.  We consider only those modes, and drop the longitudinal index in our expressions.  For simplicity we will assume that the $E_\alpha$ form a complete set of states in the $x-y$ plane.  If they do not, we can always formally add the missing states, giving them an infinite energy.

The operator which creates a photon in mode $\alpha$ is $\hat \chi_\alpha$, and can define an operator which creates a maximally focussed beam as
\begin{equation}
\hat\chi(x,y) \propto \sum_\alpha E_\alpha(x,y) \hat \chi_\alpha.
\end{equation}
By the completeness property, these obey the standard Bose commutation relations.
The Hamiltonian can be expressed as
\begin{eqnarray}
\hat H_\gamma &=& \sum_\alpha \epsilon_\alpha \hat \chi_\alpha^\dagger \chi_\alpha \\
&=& \int\!\! dr\, dr^\prime\, \hat\chi^\dagger(r) H_\chi (r,r^\prime) \hat\chi(r^\prime),\label{chieq}
\end{eqnarray}
which defines the kernel $H_\chi$.

At every point in space we envision a large number of 3-level atoms, with ground state $|g\rangle$, a short lived excited state $|p\rangle$, and a long-lived excited Rydberg state $|s\rangle$.  The cavity photons couple the states $|g\rangle$ and $|p\rangle$, and a control beam couples $|p\rangle$ and $|s\rangle$.  We introduce operators $\hat \phi_s(r)$ and $\hat \phi_p(r)$ that change the number of excitations at $r$, and work in a rotating frame.  The Hamiltonian describing these degrees of freedom are
\begin{equation}\label{hr}
\hat H_r=\int\!\! dr\left[ 
g (\hat\chi^\dagger \hat\phi_p + \hat\phi_p^\dagger \hat\chi)
+\Omega (\hat\phi_s^\dagger \hat\phi_p+\hat\phi_p^\dagger \hat\phi_s)
+\Delta \hat\phi_p^\dagger \hat\phi_p
\right]
\end{equation}
Here $\Delta=\delta-i\gamma$, where $\delta=E_p-E_s+\omega_c$ is the detuning of the p-state from resonance and $\tau=1/(2\gamma)$ is the lifetime of that state. Here $E_p$ and $E_s$ are the energies of the states in the lab frame, and $\omega_c$ is the frequency of the control laser.  The Rabi frequencies from the cavity and control lasers are $g$ and $\Omega$.  Since this Hamiltonian is strictly local, we have left off the position arguments in the operators.

It is convenient to transform to a basis which diagonalizes Eq.~(\ref{hr}), 
\begin{eqnarray}\label{psi1}
\hat\psi&\propto&\Omega \hat\chi - g \hat\phi_s\\
\hat\phi_+&\propto& g \hat\chi + E_+ \hat\phi_p +\Omega \hat\phi_s\\\label{psi3}
\hat\phi_-&\propto&
g \hat\chi + E_- \hat\phi_p +\Omega \hat\phi_s
\end{eqnarray}
These represent the long-lived dark polaritons, with energy $E_{\rm dark}=0$ and the short-lived bright polaritons with energies
\begin{equation}
E_{\pm} = \frac{\Delta \pm \sqrt{\Delta^2+4 g^2+4\Omega^2}}{2}.
\end{equation}
The relationships in Eq. (\ref{psi1}) through (\ref{psi3}) can be inverted to write
\begin{equation}
\hat \chi= \frac{\Omega}{\sqrt{\Omega^2+g^2}} \hat \psi+\eta_+  \hat \phi_+
+\eta_-  \hat \phi_-
\end{equation}
where $\eta_{\pm}= g/\sqrt{\Omega^2+g^2+|E_\pm|^2}$.
We substitute this into Eq.~(\ref{chieq}).  There are nine terms, but the
 bright polaritons are far off resonance.  Keeping only the terms involving the dark polaritons,
\begin{eqnarray}
\hat H_\gamma 
&=& \int\!\! dr\,dr^\prime\, \hat\psi^\dagger(r) H_\psi (r,r^\prime) \hat\psi(r^\prime),
\end{eqnarray}
where
\begin{equation}
H_\psi(r,r^\prime) = \frac{\Omega^2}{\Omega^2+g^2}\left[H_\chi(r,r^\prime) - \omega_0 \delta(r-r^\prime)\right] 
\end{equation}
and $\omega_0=E_p-E_g-\omega_c$ is the difference between the Rydberg excitation energy and the control frequency.  It appears due to transforming into the rotating frame.  It simply shifts the entire spectrum, and has no physical consequences.  

Our truncation is only valid if the spectrum of $H_\chi - \omega_0 {\cal I}$ is small compared to the splitting between polaritons $\sim \sqrt{\Omega^2+g^2}$.  The influence of the discarded modes can be treated perturbatively \cite{simontheory}.

The Rydberg atoms interact via a dipole-dipole interaction,
\begin{equation}
\hat H_{\rm int} =
\int\!\! dr\,dr^\prime\,V(r-r^\prime) \phi_s^\dagger (r) \phi_s^\dagger(r^\prime) \phi_s(r^\prime) \phi_s(r),
\end{equation}
where the interaction is typically modeled as 
$V(r)=C_6 r^{-6}$ -- though the exact potential is more complicated \cite{weber}.  The challenge here is that $V$ can  be much larger than any of the other scales, and hence mixes in the other polariton modes.  It is large, however, only in a small region of space.  Thus its effect on the modes is captured by a zero-range potential
\begin{equation}
\hat H_{\rm int}^{\rm eff} =
U \int\!\! dr\, \psi^\dagger (r) \psi^\dagger(r) \psi(r) \psi(r).
\end{equation}
The coefficient $U$ is found by matching the low-energy scattering phase shifts, for example by summing a set of ladder diagrams \cite{bienias}.  Due to mixing in the dark polariton states, these interactions generically contain an inelastic piece.  As we argue below, when $\delta$ is small, both of the polaritons in the collision are lost.  In a Lindblad formalism, this corresponds to the jump operator in Eq.~(\ref{jump2}).

While the full analysis of the scattering problem is tedious, the central physics is apparent by considering two polaritons at fixed locations, $r$ and $r^\prime$.  The Hilbert space is then 9-dimensional, as all three flavors of polaritons will be mixed.  We are considering the case where $\delta$ is small, so the state with two dark polaritons $|dd\rangle$ is nearly degenerate with the state containing one of each flavor of bright polaritons $|+-\rangle$.  When $V$ is small compared to $\sqrt{\Omega^2+g^2}$,  we can truncate to these two states.  In the basis $|+-\rangle, |dd\rangle$, the local Hamiltonian is then
\begin{equation}
H_{\rm local} \approx 
\left(
\begin{array}{cc}
V \Upsilon_{\pm\pm}+\Delta&\quad V \Upsilon_{d\pm}\\
V\Upsilon_{d\pm}&\quad V\Upsilon_{dd}
\end{array}
\right)
\end{equation}
with 
\begin{eqnarray}
\Upsilon_{\pm\pm}&=& \frac{1}{2} \frac{\Omega^4}{(\Omega^2+g^2)^2} \\
\Upsilon_{dd}&=&  \frac{g^4}{(\Omega^2+g^2)^2}\\
\Upsilon_{d\pm}&=& \frac{1}{\sqrt{2}} \frac{g^4}{(\Omega^2+g^2)^2}.
\end{eqnarray}
If $V=0$ the state $|dd\rangle$ is an eigenstate, with energy 0. For non-zero interactions
the state continuously connected with that one 
has an admixture of $|+-\rangle$ and its energy is found by solving a quadratic equation.  For $V\ll |\Delta|$, the mixing is weak, and the potential that the polariton feels is just the Rydberg-Rydberg coupling scaled by the overlap with the Rydberg state 
\begin{equation}
E_{\rm local}^{\rm weak}=
 V \frac{g^4}{(\Omega^2+g^2)^2}.
\end{equation}
When the interaction is strong, $V\gg |\Delta|$, the eigenstate is independent of $V$ and the effective interaction is proportional to $\Delta=\delta+i\gamma$,
\begin{equation}
E_{\rm local}^{\rm strong}=
 \Delta \frac{2 g^4}{2 g^4+\Omega^4}.
\end{equation}
Within this framework, the fact that this energy is complex indicates there is particle loss.  Since the loss comes from coupling to the state with two bright polaritons, two particles are lost.

As argued in \cite{simontheory}, when $V$ is sufficiently large, one will also have collisional loss events where only one of the polaritons are lost.  These will dominate if $\delta\gg \sqrt{\Omega^2+g^2}$.  Thus for our driven-dissipative state preparation scheme to work, we require that we are in the regime where $\delta$ is small.

\section{Multi-photon drive}
\label{appendix: multi_photon_drive}
Our protocol faces two challenges: (1) efficiently exciting from the vacuum state to the metastable 4-polariton manifold, and (2) avoiding driving the system out of the Laughlin state.  Both of these difficulties would be avoided if we could directly inject 4 photons into the system, for example with a drive term of the form
\begin{equation}
H_4=\lambda \hat a_6^\dagger \hat a_6^\dagger\hat a_6^\dagger \hat a_6^\dagger+{\rm c.c.} 
\end{equation}
Under those circumstances the Laughlin state is a true dark state (in the absence of single-particle loss).  In this Appendix we briefly explore the dynamics under this hypothetical drive, contrasting it with a two-photon drive 
\begin{equation}
H_2=\lambda \hat a_6^\dagger \hat a_6^\dagger+{\rm c.c.} 
\end{equation}
and the single-photon drive used in the main text.  In this latter case, we take $\lambda_3=\lambda_9=0$.

As in the main text, we use units where $U=1$, and take $\gamma_1=0$, but we use a significantly stronger 2-body decay rate $\gamma_2=0.1$.  The latter sets the characteristic scale for the incoherent transitions from the four-polariton manifold to the Laughlin state, and the widths of various resonances.

\begin{figure}
\includegraphics[width=\columnwidth]{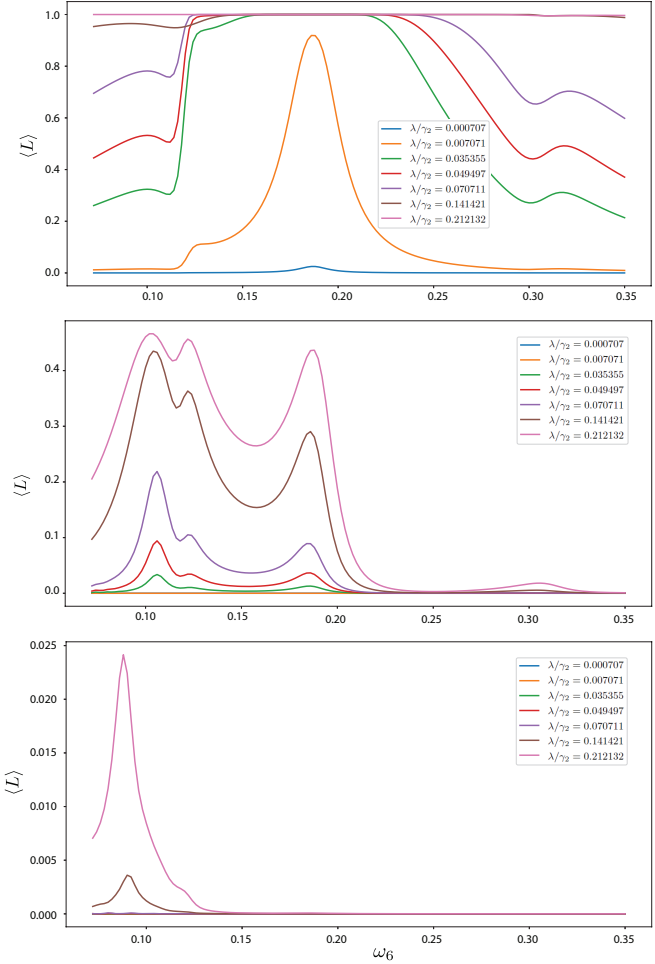}
\caption{(color online) Probability of occupying the Laughlin state at time $t=1000/\gamma_2$ for 
 four-photon (top), two-photon (middle), and single-photon (bottom) drive.  In all cases only the $m=6$ mode is driven, and  $\gamma_2 = 0.10$.  The different curves correspond to various drive amplitudes $\lambda$ relative to the two-body loss rate $\gamma_2$.}
\label{multiphoton}
\end{figure}

 There are three different four-polariton states which can be excited by these drives.  They are resonant when $\omega_6$ = 0.31, 0.19, 0.12.  Figure~\ref{multiphoton} shows the probability of being in the Laughlin state after a time $\gamma_2 t=1000$, for different drive strengths.  We choose to look at the finite-time probabilities, as in our 4-photon drive model the Laughlin state is a true dark state, and if one waits long enough the system will have a 100\% chance of occupying that state.
 
 The top panel of Fig.~\ref{multiphoton} shows results for the four-photon drive.  The dominant feature is a large peak near $\omega_6=0.19$.  This corresponds to the transition with the largest coupling matrix element.  Smaller peaks are visible near the other resonances.  As the drive amplitude increase, the maximum probability saturates at 100\%, and spreads into a plateau.
  
  The two-photon drive (middle panel) shows the same three peaks, with an additional feature closer to $\omega_6\approx 0.1$.  This corresponds to a two-photon resonance.  Generically the probabilities of occupying the Laughlin state are smaller for the two-photon drive than for the four-photon drive.  Note that interactions in the intermediate two-polariton states mix the different polariton flavors, enhancing the $\omega_6=0.12$ compared to what is seen in the top figure.
  
  The single-photon drive (bottom panel) shows essentially zero occupation of the Laughlin state, unless the drive is very strong.  This is because with only the $\lambda_6$ drive there is no way to arrange a set of resonant intermediate states.  The location of the peak that appears near $\omega_6\approx 0.09$ for very strong drive is a compromise between the final-state and intermediate-state resonances.
  
%

\end{document}